\documentclass[a4paper,twocolumn,fleqn]{article}

\usepackage{txfonts}               
\usepackage{cite}                  
\usepackage{graphicx}
\usepackage{pfr}                   
\usepackage{comment}

\begin{document}

\title{Effects of Turbulent Energy Exchange between Electrons and Ions on Global Temperature Profiles}


\author{Tetsuji~KATO\sup{1}, Hideo~SUGAMA\sup{1, 2} and Mitsuru~HONDA\sup{3}}

\affiliation{
    \sup{1} Graduate School of Frontier Science, The University of Tokyo, Kashiwa 277-8561, Japan \\
  \sup{2} National Institute for Fusion Science, Toki 509-5292, Japan \\
  \sup{3} Graduate School of Engineering, Kyoto University, Nishikyo, Kyoto 615-8530, Japan}

\date{(Received 10 January 2009 / Accepted 24 January 2009)}

\email{kato.tetsuji21@ae.k.u-tokyo.ac.jp}

\begin{abstract}
  Microscale turbulence drives not only particle and heat transport but also energy exchange between different particle species.
  Previous local gyrokinetic studies have shown that turbulent energy exchange can exceed collisional exchange in weakly collisional plasmas, and that ion temperature gradient (ITG) turbulence may hinder ion heating by alpha-heated electrons.
  In addition, it has been clarified that trapped electron mode (TEM) turbulence transfers energy from electrons to ions, thereby enhancing ion heating.
  In this work, we extend these studies by examining the impact of turbulent energy exchange on the global temperature profiles at a steady state using the one-dimensional transport solver GOTRESS. 
  For the case of DIII-D discharge \#128913~[A. E. White {\it et al.}, Phys.\ Plasmas {\bf 15}, 056116 (2008)], turbulent energy exchange has minimal influence on temperature profiles.
  However, in the case of enhanced electron heating in a DIII-D-like tokamak plasma, energy transfer from hot electrons to cold ions driven by TEM turbulence becomes comparable to, or even exceeds, the collisional contribution, leading to a significant increase in the ion temperature profile.
  For ITER Baseline and SPARC standard H-mode scenarios~[N.T. Howard {\it et al.}, Nucl.\ Fusion {\bf 65}, 016002(2024), P. Rodriguez-Fernandez {\it et al.}, J.\ Plasma\ Phys.\ {\bf 86}, 865860503(2020)], the turbulent energy exchange is largely compensated by the collisional one, producing only small effects.
  These results indicate that the impact of turbulent energy exchange on the global temperature profiles in steady‐state conditions of future fusion reactor scenarios is expected to be negligibly small, although it can become significant in situations such as plasma start-up phases, where the heating power is strongly unbalanced between electrons and ions.
\end{abstract}

\keywords{Microturbulence, Energy exchange, Energy flux, Temperature profile, Tokamak plasmas}

\DOI{10.1585/pfr.4.000}

\maketitle  

Microscale turbulence not only drives particle and heat fluxes but also induces energy exchange between electrons and ions\cite{Kato2024, Kato2025, Sugama1996, Sugama2009}.
The previous study\cite{Kato2024} has performed local gyrokinetic simulations to investigate turbulent energy exchange driven by ion temperature gradient (ITG) turbulence.
It has reported that, in weakly collisional plasmas, the turbulent contribution to the energy exchange between electrons and ions can exceed that due to Coulomb collisions.
In particular, turbulent energy exchange associated with ITG turbulence transfers energy from ions to electrons irrespective of the temperature difference between electrons and ions, and therefore may hinder the collisional energy transfer from alpha-heated electrons to ions.
Reference\cite{Kato2025} has found that trapped electron mode (TEM) turbulence transfers energy from electrons to ions, which is in the direction opposite to that of ITG turbulence.
Furthermore, it has also examined the relationship between turbulent energy exchange and the energy fluxes, deriving a simple model that allows turbulent energy exchange to be estimated with low computational cost, and confirming its validity.
In the present work, by comparing simulation results with and without the turbulent energy exchange term, we examine its impact on the steady-state global temperature profiles and investigate the necessity of incorporating turbulent energy exchange into the analysis of existing experiments and the prediction of global temperature profiles for future fusion reactors.
To predict the global temperature profiles, we employ GOTRESS\cite{Honda2018, Honda2019}, which is  a one-dimensional steady-state transport solver to solve the heat transport equations at a steady state given by,
\begin{equation}
    \label{eq: transport equation}
    0=-\frac{1}{V'}\frac{\partial}{\partial \rho}\left\{V'\left( \frac{5}{2}T_a\Gamma_a+q_a \right)\right\} +S_a,
\end{equation}
where $\rho$ is the normalized radial coordinate, and $V=V(\rho)$ is the volume enclosed by the flux surface at minor radius $\rho$.
The prime denotes the $\rho$-derivative, and $T_a=T_a(\rho)$ and $S_a=S_a(\rho)$ are the temperature and heat sources for particle species $a$ at $\rho$, respectively.
The particle and heat fluxes, $\Gamma_a=\Gamma_a(\rho)$ and $q_a=q_a(\rho)$ are written as
\begin{eqnarray}
    \label{eq:particleflux}
    & &\Gamma_a=-\left\langle|\nabla \rho|^2\right\rangle D_a \frac{\partial n_a}{\partial \rho}, \\
    \label{eq:heatflux}
    & &q_a=-n_a \left\langle|\nabla \rho|^2\right\rangle \chi_a \frac{\partial T_a}{\partial \rho},
\end{eqnarray}
where $n_a=n_a(\rho)$ is the density for particle species $a$.
The particle and heat diffusivities, $D_a=D_a(\rho)$ and $\chi_a=\chi_a(\rho)$, comprise turbulent and neoclassical contributions: $D_a=D_a^{\rm turb}+D_a^{\rm neo}$, $\chi_a=\chi_a^{\rm turb}+\chi_a^{\rm neo}$.

\begin{table}[b]
\centering
    \caption{Parameters for each device case}
\begin{tabular}{lccc} \hline
                    & DIII-D&    ITER& SPARC \\ \hline
    Major radius $R_0~[{\rm m}]$&   1.66&    6.2 &   1.85 \\
    Minor radius $r_0~[{\rm m}]$&   0.67&    2.0 &   0.57    \\
    Toroidal field $B_0~[{\rm T}]$&  2.1&   5.3& 12.2\\
    Plasma current $I_p~[{\rm MA}]$& 1.0& 15.0& 8.7    \\
    \hline
    \label{tab:Devices}
\end{tabular}
\end{table}

The total cumulative power $P_a^{\rm total}=P_a^{\rm total}(\rho)$ inside the volume $V(\rho)$ is calculated by
\begin{equation}
    \label{eq:integratedpower}
    \hspace{-1.0cm} P_a^{\rm total}(\rho)=\int_{\rho}S_aV'd\rho=P_a^{\rm heat}(\rho)+P_a^{\rm coll}(\rho)+P_a^{\rm turb}(\rho),
\end{equation}
where the cumulative power for energy exchange between electrons and ions via Coulomb collision and microturbulence, $P_a^{\rm coll}(\rho)=\int_{\rho}S_a^{\rm coll}V'd\rho$ and $P_a^{\rm turb}(\rho)=\int_{\rho}S_a^{\rm turb}V'd\rho$ for $a=e, i$, respectively.
Here, for convenience, the heat source $S_a$ is taken to include not only auxiliary heating, alpha heating, and ohmic heating, but also radiation loss for electrons as negative contributions.
The turbulent energy transfer from electrons to ions $S_i^{\rm turb}(=-S_e^{\rm turb})$ can be modeled  as\cite{Kato2025}
\begin{equation}
    \label{eq:TEE}
    S_i^{\rm turb}=\frac{\mathcal{E}^{\rm turb}_{e}-\mathcal{E}^{\rm turb}_{i}}{2R_0},
\end{equation}
where $\mathcal{E}^{\rm turb}_{a}=q_a^{\rm turb}+5T_a\Gamma_a^{\rm turb}/2$, $\Gamma^{\rm turb}_{a}=D_a^{\rm turb}\Gamma_a/D_a$, $Q^{\rm turb}_{a}=\chi_a^{\rm turb}Q_a/\chi_a$ for $a=e, i$, and $R_0$ is the major radius.
In this study, we incorporate the effect of turbulent energy exchange into GOTRESS using Eq.~(\ref{eq:TEE}) and investigate several tokamak cases as shown in Tab.~\ref{tab:Devices}.
Here, for simplicity, we assume a tokamak geometry with magnetic flux surfaces that have concentric circular cross sections, noting that in Eq.~(\ref{eq:TEE}) only the major radius $R_0$ and the turbulent energy fluxes $\mathcal{E}_a^{\rm turb}$ ($a=e,i$) are used to evaluate the turbulent energy exchange.
The density profile is fixed, and the temperature profile is solved for.
The current density profile is assumed to be proportional to $(1 - \rho^{2})^{2}$.

\begin{figure}[tb]
    \includegraphics[keepaspectratio, width=0.48\textwidth]{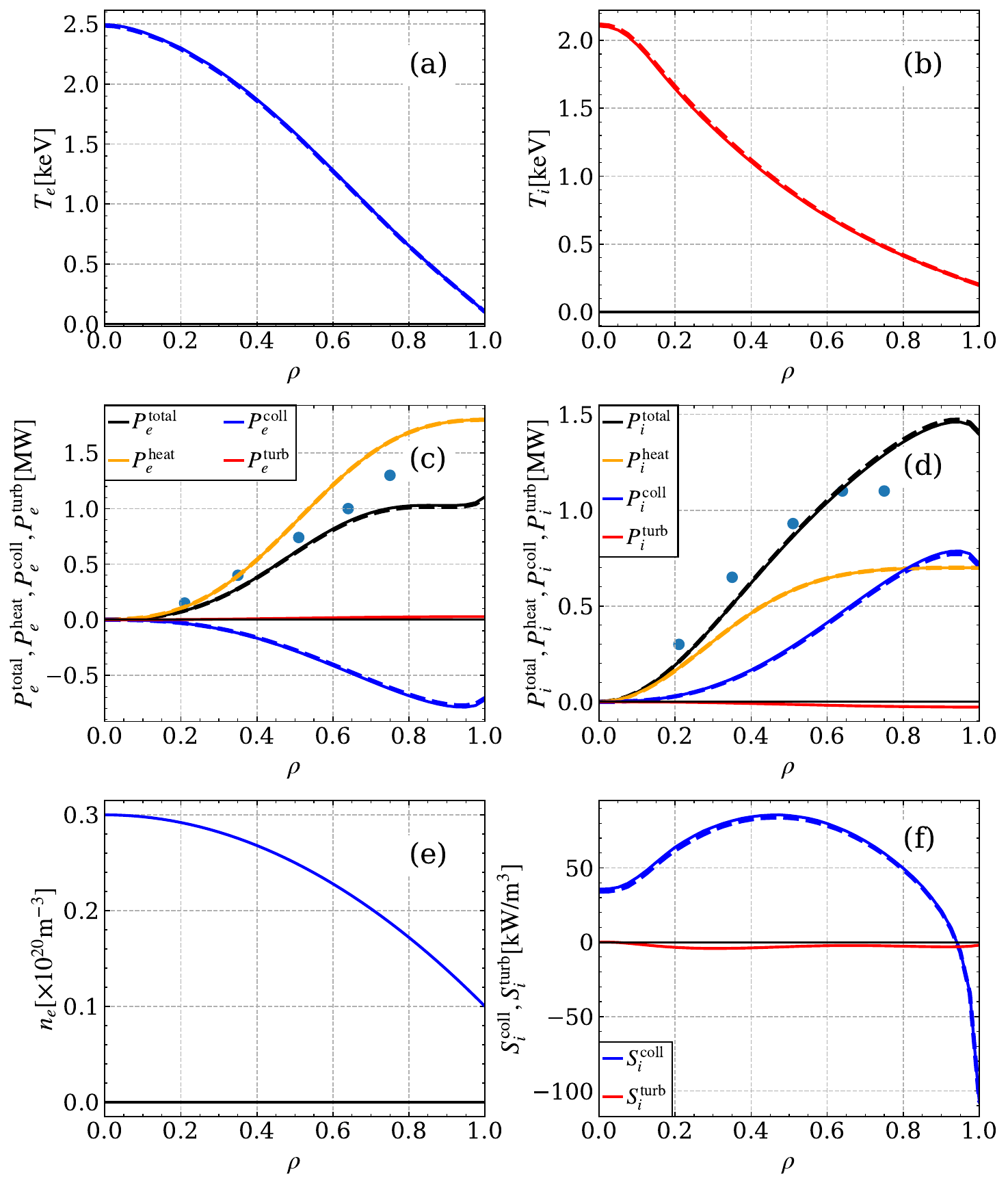}
    \caption{Simulation results for conditions similar to the DIII-D discharge~\#128913 with the parameters $(c_e, c_i)=(3.0, 6.0)$ and heating powers $(P_e^{\rm heat}, P^{\rm heat}_i)=(1.8, 0.7)$ [MW]. 
    Figures~\ref{fig:DIII-D128913}(a–f) present, respectively, the global electron and ion temperature profiles, the cumulative electron and ion power, the density profile, and the collisional and turbulent energy exchanges.
    The circular markers in Figs.~\ref{fig:DIII-D128913}(c) and (d) indicate the energy transport across a flux surface at each minor radius reported in Ref.~\cite{Holland}.
    Solid and dashed lines represent results with and without the turbulent energy exchange described in Eq.~(\ref{eq:TEE}), respectively.
    }
    \label{fig:DIII-D128913}
\end{figure}

In this study, we express the turbulent particle and heat diffusivities as
$D^{\rm turb}_a = 0.2\chi^{\rm turb}_a$ (see e.g.~\cite{Takenaga}) and $\chi_a^{\mathrm{turb}} = c_a \chi_{a,\mathrm{BgB}}^{\mathrm{turb}}$, where $\chi_{a,\mathrm{BgB}}^{\mathrm{turb}}$ denotes the value obtained from the BgB model.
The BgB model enables computationally efficient evaluations of microscale turbulence \cite{Vlad, Erba}.
The tuning parameters $c_a$ for $a=e, i$ are adjusted so that the resulting temperature profiles and energy balance reproduce the experimental or reference data.
Figure~\ref{fig:DIII-D128913} shows simulation results for conditions similar to the DIII-D discharge \#128913, representing an ITG-dominant case\cite{Kato2024, White, Candy, Holland}.
The circular markers in Figs.~\ref{fig:DIII-D128913}(c) and (d) indicate the energy transport across a flux surface at each minor radius reported in Ref~\cite{Holland}.
As shown in Fig.~\ref{fig:DIII-D128913}(f), the turbulent energy transfer contributes in the opposite direction to the collisional one.
However, for the entire volume, $P_i^{\mathrm{turb}}(\rho=1)(=-P^{\rm turb}_e(\rho=1))$ is only 2.4\% and 1.9\% of $P_e^{\mathrm{total}}(\rho=1)$ and $P_i^{\mathrm{total}}(\rho=1)$, respectively, and the resulting change in the temperature profiles is negligible as shown in Figs.~\ref{fig:DIII-D128913}(a) and (b).
This result is consistent with previous studies~\cite{Kato2024, Candy}, indicating that in this case the turbulent energy exchange can be safely neglected.

\begin{figure}[tb]
    \includegraphics[keepaspectratio, width=0.49\textwidth]{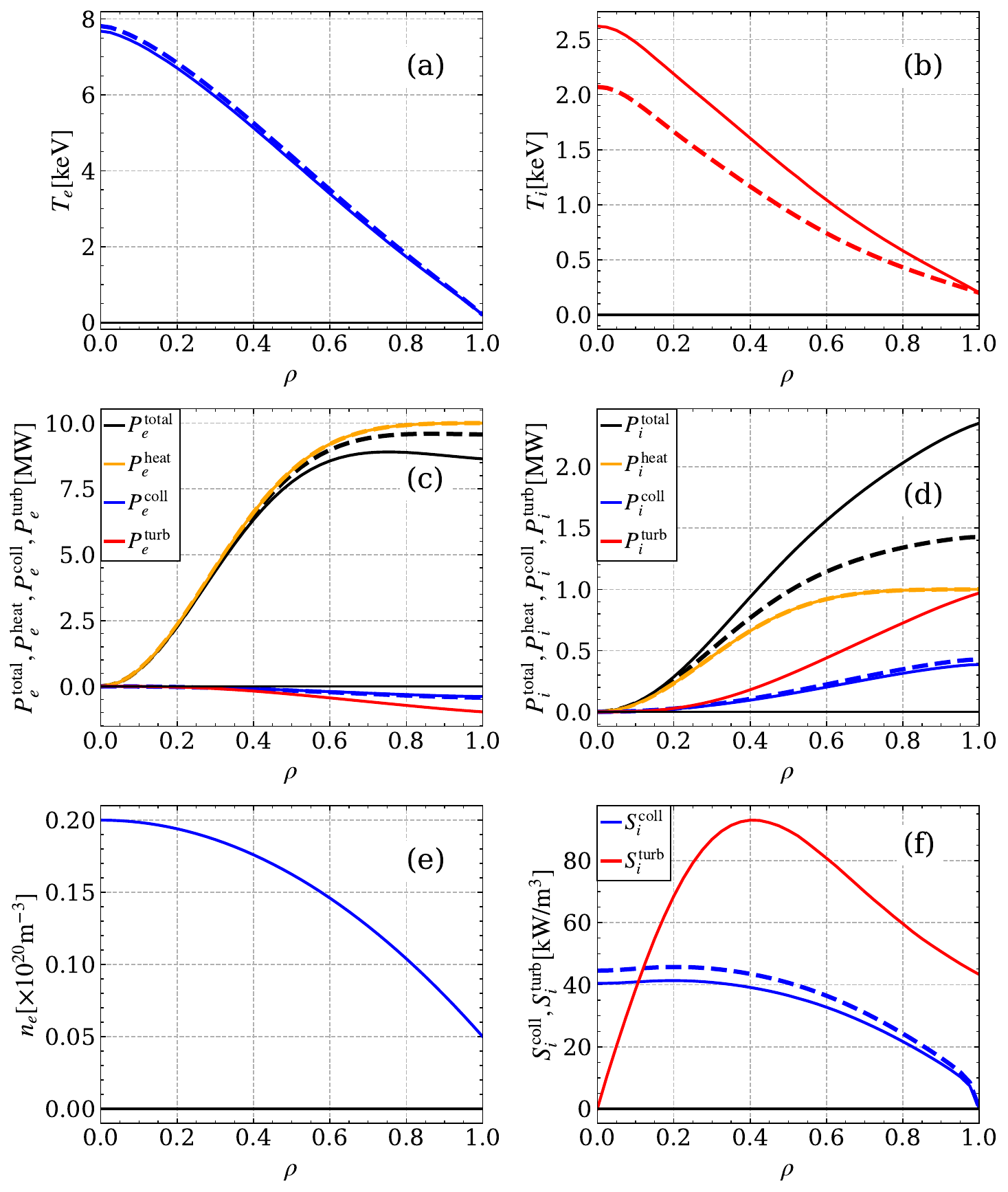}
    \caption{Simulation results for DIII-D-like tokamak with the parameters $(c_e, c_i)=(5.0, 3.0)$ and heating powers $(P_e^{\rm heat}, P^{\rm heat}_i)=(10, 1.0)$ [MW].
    Figures~\ref{fig:DIII-D_caseB}~(a–f) present the same quantities as those described in the caption of Fig.\ref{fig:DIII-D128913}.
    }
    \label{fig:DIII-D_caseB}
\end{figure}

%
Next, we consider the DIII-D–like tokamak with significantly enhanced electron heating ($P_{e}^{\mathrm{heat}} = 10\;\mathrm{MW},\; P_{i}^{\mathrm{heat}} = 1.0\;\mathrm{MW}$), which  corresponds to a TEM-dominant case, as shown in Fig.~\ref{fig:DIII-D_caseB}.
Here, the parameters $c_a$ for $a=e, i$ are adjusted based on results from the Multi-Mode Model (MMM)\cite{Rafiq}, which is used to compute turbulent transport. 
The MMM is based on the Weiland model\cite{Weiland} and can describe turbulence driven by microinstabilities such as ITG and TEM.
This module has been widely used to predict density and temperature profiles and to analyze the underlying transport physics\cite{Pankin, Kinsey}.
Although the MMM module can provide appropriate turbulent transport levels, it tends to exhibit poorer convergence in the temperature profile calculations for the case of ITG-dominant cases than for TEM-dominant cases.
In particular, the turbulent heat diffusivity for ions often becomes sharply small near the magnetic axis, making profile prediction difficult. 
Therefore, for this case, we tuned the BgB model using the MMM results to obtain smooth, continuous turbulent diffusivities.

\begin{figure}[tb]
    \includegraphics[keepaspectratio, width=0.48\textwidth]{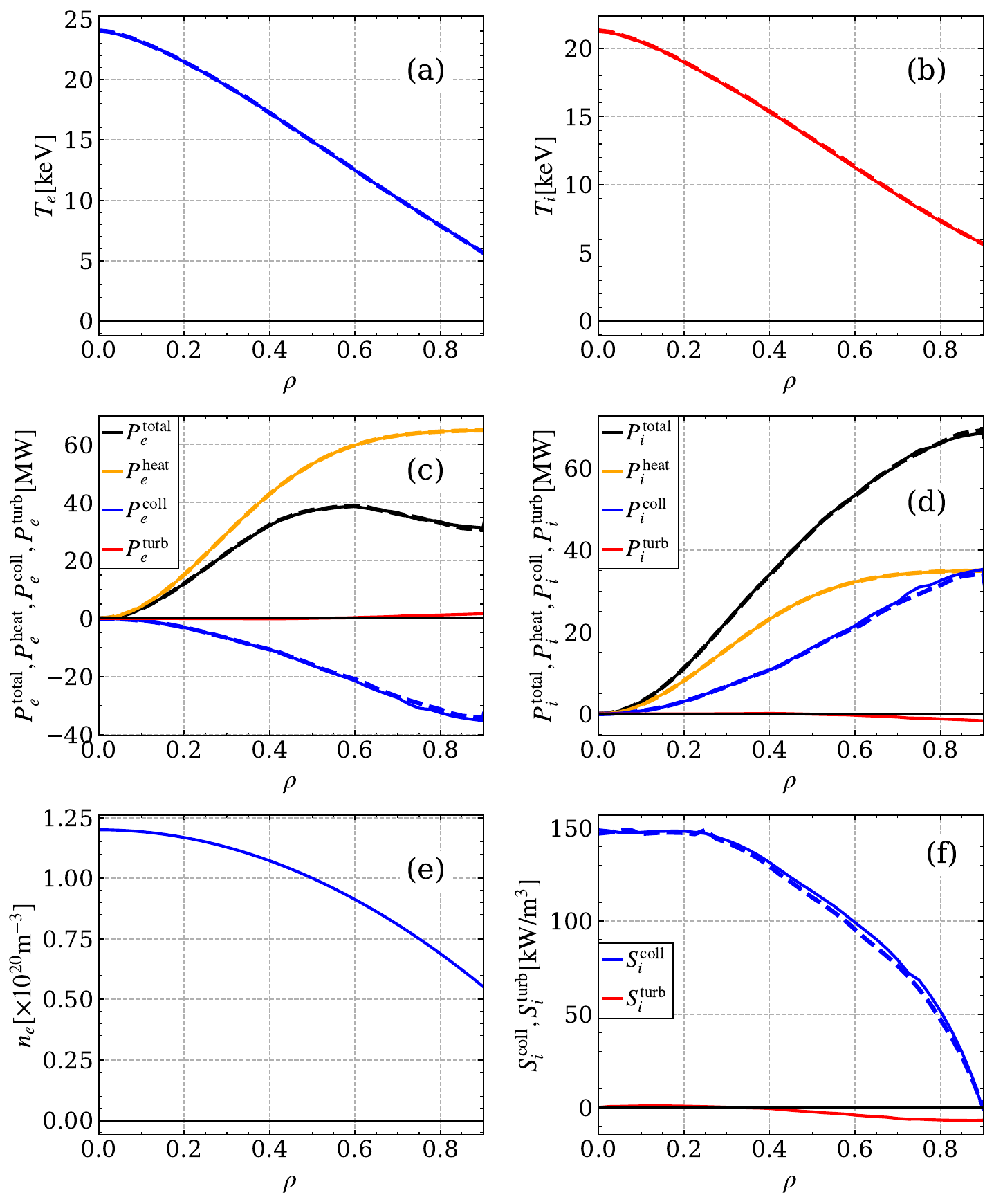}
    \caption{Simulation results for conditions similar to the ITER Baseline scenario\cite{Howard} with the parameters $(c_e, c_i)=(2.0, 7.0)$ and heating powers $(P_e^{\rm heat}, P^{\rm heat}_i)=(65, 35)$ [MW].
    Figures~\ref{fig:ITER}~(a–f) present the same quantities as those described in the caption of Fig.\ref{fig:DIII-D128913}.}
    \label{fig:ITER}
\end{figure}

%
In this high electron heating case, the electron energy flux significantly exceeds the ion energy flux.
Consequently, Fig.~\ref{fig:DIII-D_caseB}(f) shows that energy is transferred by microturbulence from electrons to ions, in the same direction as the collisional energy exchange.
Moreover, the magnitude of the turbulent energy exchange exceeds that of the collisional one by more than a factor of two, indicating that turbulence becomes the dominant contribution to the energy exchange between electrons and ions. 
For the entire device, as shown in Figs.~\ref{fig:DIII-D_caseB}(c) and (d), the net energy transfer from electrons to ions increases by approximately 1~MW due to turbulent energy exchange, which is 41~\% of $P_i^{\rm total}(\rho=1)$ for the result with turbulent energy exchange.
This increase of energy transfer significantly modifies temperature profiles, and the temperature difference at magnetic axis for electrons and ions, $\Delta T_e$ and $\Delta T_i$, are -0.13~keV and 0.55~keV, respectively.
These results demonstrate that turbulent energy exchange can play a crucial role when the heating power is strongly unbalanced or when there is a large disparity between electron and ion energy fluxes.

\begin{figure}[tb]
    \includegraphics[keepaspectratio, width=0.48\textwidth]{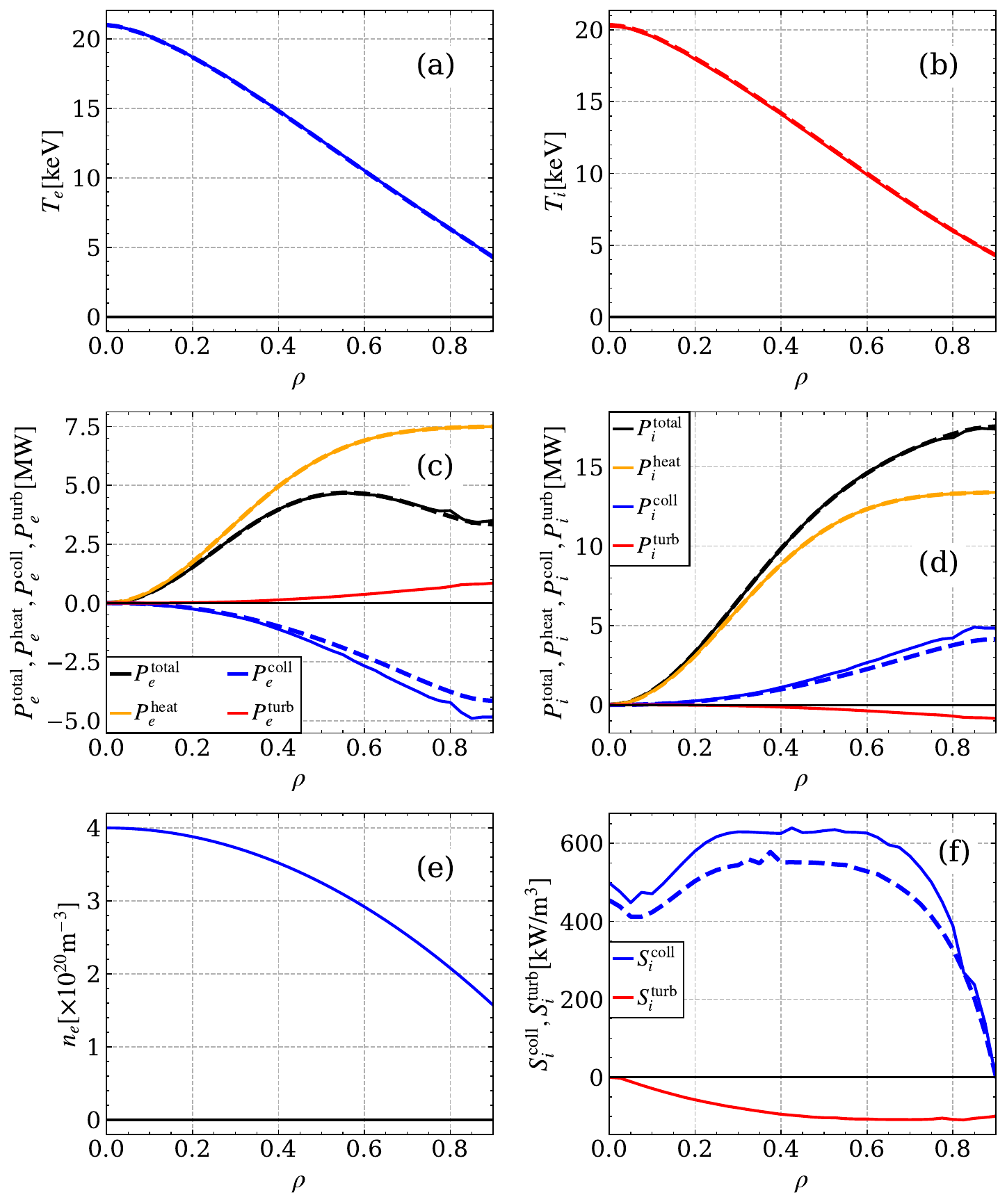}
    \caption{Simulation results for conditions similar to the SPARC standard H-mode scenario\cite{Fernandez} with the parameters $(c_e, c_i)=(0.5, 3.5)$ and heating powers $(P_e^{\rm heat}, P^{\rm heat}_i)=(7.5, 13.4)$ [MW].
    Figures~\ref{fig:SPARC}~(a–f) present the same quantities as those described in the caption of Fig.\ref{fig:DIII-D128913}.}
    \label{fig:SPARC}
\end{figure}

%
We now examine reactor-relevant cases using ITER- and SPARC-like conditions.
In both cases, the parameters $c_a$ for $a=e,i$ are adjusted using simulation results reported in Refs.~\cite{Howard, Fernandez}. 
Here, the boundary is set at $\rho = 0.9$ to exclude the complex physics in the H-mode pedestal region.
In addition, while the reference studies consider deuterium-tritium plasmas, this work treats ions as a single species.
Figures~\ref{fig:ITER} and \ref{fig:SPARC} show the results under the ITER Baseline and SPARC standard H-mode scenarios, respectively.
Collisional energy transfer from alpha-heated electrons to ions is expected to be essential to maintain a high ion temperature.
In addition, ITG turbulence is dominant, and the turbulent energy flux for ions $\mathcal{E}^{\rm turb}_i$ exceeds the flux for electrons $\mathcal{E}^{\rm turb}_e$.
Accordingly, the turbulence transfers energy from ions to electrons($S^{\rm turb}_i<0$), opposite to the collisional one($S^{\rm coll}_i>0$).
The model of turbulent energy exchange described in Eq.~(\ref{eq:TEE}) indicates that the exchange depends on the energy fluxes, which represent the amount of energy crossing a magnetic flux surface per unit area.
Although the total cumulative power is larger in ITER than in SPARC, the turbulent energy exchange $S^{\rm turb}_a (a=e, i)$ is larger in SPARC because its smaller device size leads to larger energy fluxes.
For the plasmas inside $\rho=0.9$, the cumulative power for turbulent energy transfer from ions to electrons $P^{\rm turb}_e(\rho=0.9)(=-P^{\rm turb}_i(\rho=0.9))$ is 1.7~MW  in ITER and 0.85~MW in SPARC.
This effect, however, is largely compensated by a slight increase in the temperature difference across the minor radius, which enhances the collisional energy exchange $S_a^{\rm coll}(a=e, i)$.
Then, the cumulative power for the collisional energy exchange $P^{\rm coll}_i(\rho=0.9)$ increases by 1.0~MW in ITER and 0.69~MW in SPARC.
Consequently, the net energy exchange between electrons and ions remains nearly unchanged.
These results indicate that, for high-density plasmas relevant to future fusion reactors, the modification of the temperature profiles due to turbulent energy exchange is relatively small. 
Nevertheless, in smaller devices such as SPARC, the turbulent contribution in the energy exchange can become relatively more significant.
In this study, the impact of turbulent energy exchange on steady-state temperature profiles is investigated.
As long as the heating powers for ions and electrons do not differ drastically (e.g., by less than a factor of two), the contribution of turbulent energy exchange remains small.
However, when large disparities in the heating or energy fluxes of electrons and ions occur, turbulent energy exchange becomes comparable to, or even exceeds, the collisional contribution, substantially altering the temperature profiles.
These results highlight scenarios in which turbulent energy exchange must be incorporated for accurate reactor-scale predictions.
It is worthwhile to investigate the impact of turbulent energy exchange in plasma start-up scenarios, where electron heating is dominated by Ohmic heating and ions are heated solely through energy transfer from electrons.\cite{Hoppe}.
As a future task, it remains to validate the results using the MMM-calibrated BgB model by performing gyrokinetic simulations over the entire minor radius.
In addition, since the plasma start-up phases are inherently non-steady-state, further investigation is required to properly address this regime.

The present study is supported in part by the JSPS Grants-in-Aid for Scientific Research Grant No.~24K07000 and in part by the NIFS Collaborative Research Program NIFS23KIPT009. 
This work is also supported by JST SPRING, Grant Number JPMJSP2108.
Simulations in this work were performed on “Plasma Simulator” with the support and under the auspices of the NIFS Collaboration Research program (NIFS24KISM007).

\end{document}